\documentclass[onecolumn,authoryear]{els-mrw} 

\usepackage{amsmath,amssymb,amsfonts,amsthm,makeidx,graphicx}
\usepackage{txfonts}
\usepackage{helvet}
\usepackage{xcolor}
\usepackage{hyperref}
\usepackage{subcaption}
\usepackage{journalnames}

\renewcommand{\vec}[1]{\boldsymbol{#1}}

\newcommand{\ppdot}{$P\dot P$\ }

\def\cm2{cm$^{-2}$}

\def\deg{$^{\circ}$}

\def\3xmm{3XMM\,J1852$+$0033}

\begin{document}

\chapter{Magnetars}\label{chapmagnetar}

\author[1,2]{Nanda Rea}
\author[1,2]{Davide De Grandis}
%\author[2,3,4]{}

\address[1]{\orgname{Institute of Space Sciences (ICE, CSIC)}, \orgaddress{Campus UAB, Carrer de Can Magrans s/n, 08193, Barcelona, Spain}}
\address[2]{\orgname{Institut d’Estudis Espacials de Catalunya (IEEC), \orgaddress{Esteve Terradas 1, RDIT Building, 08860, Castelldefels, Spain}}}
%\address[3]{\orgname{}, \orgaddress{}}
%\address[4]{\orgname{}, \orgaddress{}}

\articletag{Chapter Article tagline: update of previous edition,, reprint..}

\maketitle

%\tableofcontents

\begin{abstract}[Abstract]
Magnetars are the most magnetic objects in the Universe, serving as unique laboratories to test physics under extreme magnetic conditions that cannot be replicated on Earth. They were discovered in the late 1970s through their powerful X-ray flares, and were subsequently identified as neutron stars characterized by steady and transient emission across the radio, infrared, optical, X-ray, and gamma-ray bands. In this chapter, we summarize the current state of our experimental and theoretical knowledge on magnetars, as well as briefly discussing their relationship with supernovae, gamma-ray bursts, fast radio bursts, and the transient multi-band sky at large. 
\end{abstract}

%HERE: Keyword -- 5-10 words that embody the key topics in the chapter. What terms would someone put into a search engine if they were looking for a chapter like this? ( https://astrothesaurus.org/, please use this site for the keywords to be included in the chapter)
Keyword -- Compact objects (288);
Magnetars (992);
Magnetic fields (994);
Magnetic stars (995);
%Millisecond pulsars (1062);
Neutron stars (1108);
%Optical pulsars(1173);
Pulsars (1306);
Radio pulsars (1353);
%Rotation powered pulsars (1408).
%Magnetic poles (2319).
Soft gamma-ray repeaters (1471);
%Irregular variable stars (865);
High energy astrophysics (739);
Radio transient sources (2008).
 
%\begin{glossary}[Glossary]
%A dictionary-style definition of any unusual or key terms used in your article. (This section can be combined into a single section with Nomenclature)
%\end{glossary}

\begin{glossary}[Nomenclature]
%System of abbreviations/terms/symbols used in the specific field of study/community. List and define (This section can be combined into a single section with Glossary).\\
%\ali{maybe we can do only the nomenclature section}\\
%\ali{PSR, RPP, RRAT, XDIN, CCO}
%\fra{P, Pdot, tau, Bdip}
\begin{tabular}{@{}lp{34pc}@{}}
AXP & anomalous X-ray pulsar\\
(e)MHD & (electron) magneto-hydrodynamics\\
FRB & fast radio burst\\
GRB & gamma-ray burst \\
NS & neutron star \\
QED & quantum electrodynamics \\
SGR & soft gamma repeater\\
XDINS & X-ray dim isolated NS

\end{tabular}
\end{glossary}

%HERE: Key points (see example chapter) -- A short bulleted list of key points or objectives for the chapter
\section*{Key points}
\begin{itemize}
    \item Magnetars are a class of isolated neutron stars characterized by an ultra-strong magnetic field.
    \item Magnetars were discovered as sources showing flaring X-ray and $\gamma$-ray activity (SGRs) and pulsars with anomalously large X-ray luminosity (AXPs).
    \item Magnetars emit in different energy bands, from radio to $\gamma$-rays.
    \item They exhibit a variety of transient, violent phenomena at high energy.
    \item They can be used as a laboratory for understanding physics at ultra-high magnetization.
    \item Recently, magnetars flaring emission or formation have been proposed to power bright extra-Galactic transients.
 \end{itemize}
 
\section{Introduction}
\label{mag:intro}

The hint for magnetars, a type of neutron star (NS) with extremely powerful magnetic fields dates back to the late 1970s when sudden bursts X/$\gamma$-rays were discovered \citep{1986Natur.322..152L}, which did not fit the patterns of other celestial sources, i.e.\ they were repeating, not single events like typical Gamma-Ray Bursts (GRBs). In particular, the 1979 detection of a Giant flare \citep[see also section \ref{sec:transients}]{1979SvAL....5..163M} from SGR\,0525$-$66 in the Large Magellanic Cloud was a landmark event to establish the class of sources known as Soft Gamma Repeaters (SGRs), characterized by this very energetic bursting activity. 

\begin{figure}
    \centering
    \includegraphics[width=0.95\linewidth]{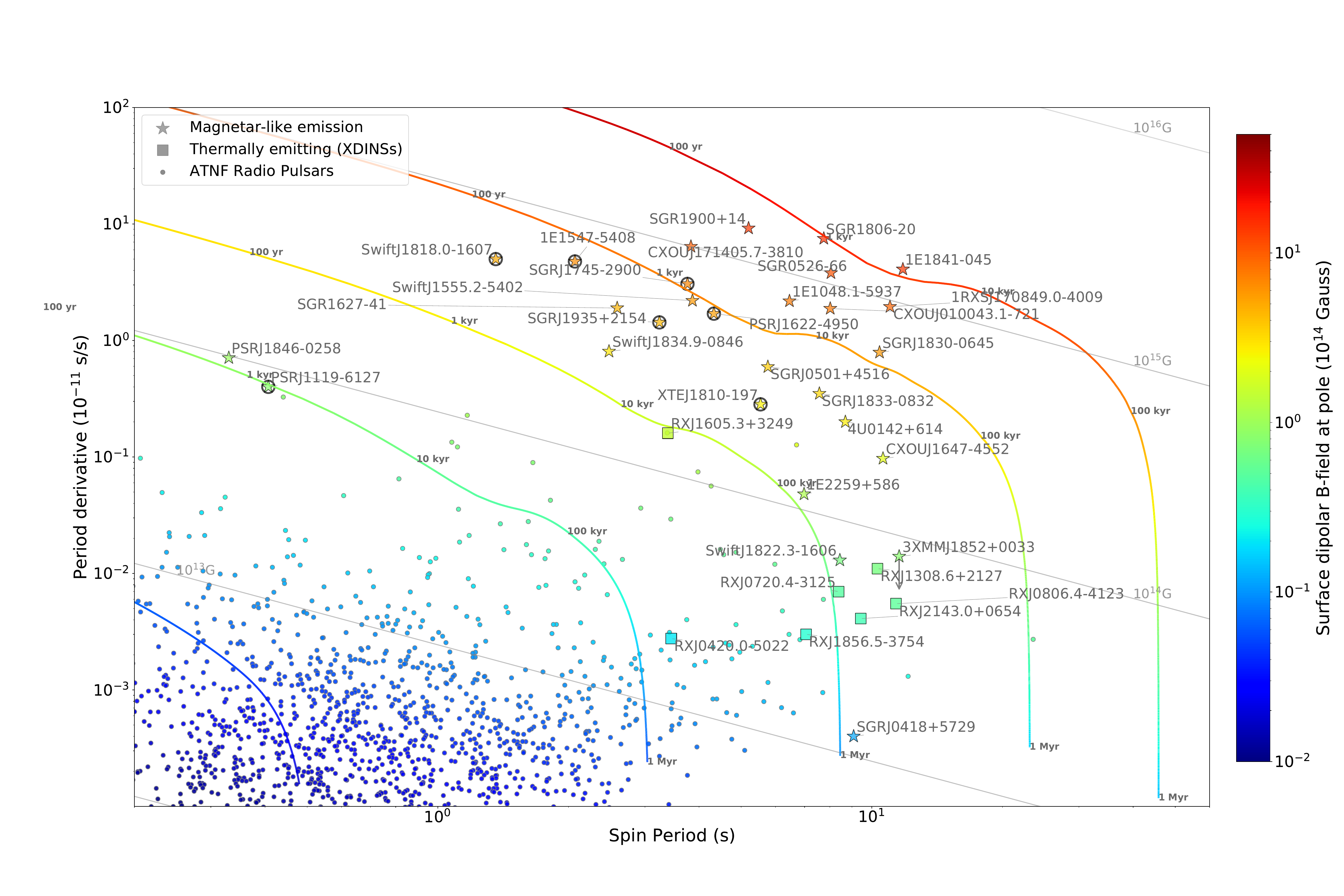}
    \caption{Period-Period Derivative diagram for different pulsar classes. The color code represent the dipolar magnetic field, and magneto-thermal evolutionary lines are reported for a range of initial magnetic fields. Radio-detected magnetars are marked with a black circle.}
    \label{fig:ppdot}
\end{figure}

At that time pulsars were already well known, and their emission mechanism had been linked to the presence of a strongly magnetized, rotating body \citep{1968Natur.218..731G}. Indeed, the comparison of pulsar radio timing data to the spin down expected from a rotating magnetic dipole, showed that the bulk of the radio pulsar population was endowed with magnetic fields of $\approx10^{12}$\,G (see figure \ref{fig:ppdot}). As large as this value might be for terrestrial standards, these new discoveries following the advent of space-based high-energy instruments, pointed to the existence of even more extreme magnetic sources.

On the other hand, by the 1990s hints of ultra-strong fields were found in a class of persistent X-ray sources with atypically large X-ray luminosity ($\approx10^{34}-10^{36}\,$erg/s) and whose spin period and spin-down were those of a dipole with $B_d\approx10^{14}\,$G, dubbed Anomalous X-ray Pulsars \citep[AXPs,][]{mereghetti95}. 
These two classes of SGRs and AXPs were unified by \citet{duncan92, 1993AIPC..280.1085T} under the \emph{magnetar} paradigm. Later on, the measure of the X-ray quiescent state of SGRs with distinctive spin periods \citep{Kouveliotou1998,Kouveliotou1999}, and the detection of bursting activity from some known AXPs \citep{Gavriil2002}, further rendered the distinction obsolete, well establishing magnetars as a proper class of pulsars.

Thus, we can define a magnetar as a pulsar whose emission, both during transient, violent episodes and its persistent state, is powered by a strong magnetic field. Their spin periods range between $\approx$0.3-12\,s with relatively large period derivatives ($\dot{P}\sim10^{-13}-10^{-9}$s\,s$^{-1}$; see figure \ref{fig:ppdot}).
This is often reflected in a high value of their magnetic fields as obtained from timing measurements, $B_d\gtrsim10^{14}\,$G. This field is often larger than the \emph{critical quantum field}, i.e.\ the one at which the cyclotron energy $\hbar\omega_B=\hbar eB/m_ec$ of an electron is equal to its rest mass energy $m_ec^2$, $B_Q= m_e^2 c^3/e\hbar \simeq4.414\times10^{13}$\,G. These large fields are making magnetars ideal laboratories for the study of quantum electro-dynamics and plasma physics under strong field regimes.
Nevertheless, while the spin-down field is generally a good proxy for the actual field strength, and indeed most sources have $B_d\gtrsim B_Q$, see figure \ref{fig:ppdot}, it must be born in mind that $B_d$ is an estimate of the dipolar part of the poloidal field component, and most of the magnetic energy may well be stored in higher multipoles and/or the toroidal component. An outstanding illustration of this fact comes from the discovery of the low field magnetars, which despite having a spin-down field more akin to normal pulsars, they underwent magnetar-like bursting episodes \citep[see also section \ref{sec:transients}]{rea10, 2012ApJ...754...27R}. 

At the time of writing, 29 sources have been unequivocally identified as magnetars, all in our Galaxy or in the Magellanic Clouds (see figure \ref{mag:discoveries}). Extensive reviews about magnetars, both from observations and theoretical perspectives, can be found in \cite{2015RPPh...78k6901T,2017ARA&A..55..261K,2018smfu.book..321M,2021ASSL..461...97E}.

%%%%%%%%%%%%%%%%%%%%%%%%%%%%%%%%%%%%%
%%%%%%%%%%%%%%%%%%%%%%%%%%%%%%%%%%%%%

%%%%%%%% TABLE %%%%%%%%%%%%%%%%%%%%%%
\begin{table*}
\caption{Timing properties and timing-inferred parameters, and physical properties of the current sample of known magnetars.}
\begin{threeparttable}
\resizebox{1\columnwidth}{!}{
\begin{tabular}{rccccccccl}
\toprule
Source  & $P$& $\dot{P}$ & $B_{\rm{p, dip}}$$^{\rm a}$	&$\dot{E}_{\rm {rot}}^{\rm b}$ 	& $\tau^{\rm c}_c$	& $D$	& Association& Bands$^{\rm{d}}$ & Reference	 \\
& (s)		& (10$^{-11}$~s~s$^{-1}$)		& (10$^{14}$~G)	& (erg~s$^{-1}$)		& (kyr)		& (kpc)	      &	&	\\
\midrule
SGR\,0526$-$66$^\dagger$  &  8.05  &  3.8  &  11.0  &  $2.9 \times 10^{33}$  &  3.4   &  49.7    & SNR N49 (LMC)  & X             & \citet{olausen14} \\ 
SGR\,1900$+$14$^\dagger$  &  5.20  &  9.2  &  14.0  &  $2.6 \times 10^{34}$  &  0.9   &  12.5    & cluster        & $\gamma$ X O& \citet{olausen14} \\ 
SGR\,1806$-$20$^\dagger$  & 7.55   & 76.95 &  49    &  $7.0 \times 10^{34}$   & 0.2   &  8.7     & cluster W31    & $\gamma$ X O& \citet{2015ApJ...809..165Y}     \\
1E\,2259$+$586              & 6.98   &  0.048  &  1.2 &  $1.3 \times 10^{32}$  &  230   &  3.2     & SNR CTB109     & $\gamma$ X O I& \citet{dib14}  \\
1E\,1048.1$-$5937           &  6.46  &  2.18  &  7.6  &  $3.2 \times 10^{33}$  &  4.7   &  9       &                & $\gamma$ X O  & \citet{dib14} \\   
4U\,0142$+$614              &  8.69  &  0.20  &  2.7  &  $1.3 \times 10^{32}$  &  69    &  3.6     &                & $\gamma$ X O I&  \citet{olausen14}\\
1E\,1841$-$045              &  11.79 &  4.09  &  13.8 &  $9.9 \times 10^{33}$  &  4.6   &  8.5     & SNR Kes73      & $\gamma$ X O  & \citet{olausen14} \\
1RXS\,J170849.0$-$4009      &  11.01 &  1.95  &  9.3  &  $5.8 \times 10^{32}$  &  9.1   &  3.8     &                & $\gamma$ X O& \citet{olausen14} \\
 SGR\,1627$-$41$^{\rm e}$   &  2.59  &  1.9  &  4.5   &  $4.3 \times 10^{34}$  &  2     &  11      & SNR G337.0-0.1 & X I           & \citet{esposito09}  \\ 
XTE\,J1810$-$197            &  5.54  &  0.283  &  2.6 &  $6.7 \times 10^{32}$  &  31    &  3.5     &                & X O I R       & \citet{camilo16} \\  
CXOU\,J010043.1$-$721       &  8.02  &  1.88  &  7.9  &  $1.4 \times 10^{33}$  &  6.8   &  62.4    & SMC            & X O           & \citet{olausen14} \\
 CXOU\,J164710.2$-$455216   &  10.61 &  0.097  &  2.1 &  $3.2 \times 10^{31}$  &  173   &  4       & cluster Wd1    &   X           & \citet{2014MNRAS.441.1305R} \\  
PSR\,J1846$-$0258           &  0.33  &  0.71  &  0.98 &  $8.1 \times 10^{36}$  &  0.7   &  6.0     &                & $\gamma$ X  & \citet{vigano13}) \\ 
1E\,1547$-$5408             &  2.07  &  4.77  &  6.4  &  $2.1 \times 10^{35}$  &  0.7   &  4.5     &SNR G327.24-0.13& $\gamma$ X O R& \citet{2012ApJ...748....3D}\\  
SGR\,0501$+$4516            &  5.76  &  0.594  &  3.7 &  $1.2 \times 10^{33}$  &  15    &  1.5     & SNR G160.9+2.6 & $\gamma$ X O& \citet{2014MNRAS.438.3291C}  \\   
SGR\,0418$+$5729            &  9.08  & 0.0004 &  0.1  &  $2.1 \times 10^{29}$  &  $\sim 36000$ & 2 &                & X           & \citet{rea13}\\   
SGR\,1833$-$0832            &  7.57  &  0.35  &  3.3  &  $3.2 \times 10^{32}$  &  34 & 10$^{\rm f}$&                & X           & \citet{esposito11} \\   
PSR\,J1622$-$4950           & 4.33   & 1.7      & 2.7 &  $8.3 \times10^{33}$   &  4.0   &9         & SNR G333.9+0.0 & X R         & \citet{levin10}\\ 
CXOU\,J171405.7$-$3810      &  3.83  &  6.40  &  10.0 &  $4.5 \times 10^{34}$  &  0.95  &  13.2    & SNR CTB37B     & X           & \citet{olausen14} \\
Swift\,J1834.9$-$0846       &  2.48  &  0.806  &  2.9 &  $2.1 \times 10^{34}$  &  5     &  4.2     &              & X            & \citet{esposito13}  \\   
Swift\,J1822.3$-$1606       &  8.44  &  0.013  &  0.7 &  $8.4 \times 10^{30}$  &  1030  &  1.6     &                & X           & \citet{rodriguez16}\\   
SGR\,1745$-$2900            &  3.76  &  3.06  &  6.9  &  $2.2 \times 10^{34}$  &  1.9   &  8.3     & Galactic centre& $\gamma$ X R  & \citet{cotizelati17} \\   
3XMM\,J185246.6+003317      & 11.56  & $<0.014$ &$<0.41 $ & $<3.6\times 10^{30}$          & $>1300$& 7.1      &     & X            & \citet{2014ApJ...781L..17R}\\
SGR\,1935$+$2154            &  3.24  &  1.43  &  4.4  &  $1.6 \times 10^{34}$  &  3.6   &  9       &                & $\gamma$ X R  & \citet{2016MNRAS.457.3448I} \\  
PSR\,J1119$-$6127           &  0.41  &  0.4  &  0.82  &  $2.5 \times 10^{36}$  &  1.6   &  8.4     & SNR G292.2$-$0.5 & X R         & \citet{vigano13} \\  
1E\,161348$-$5055           &  24030 & $<$70 & $<$2600  & $<2 \times 10^{24}$   &  $>$ 540&  3.3    & SNR RCW 103   & X             & \citet{rea16}\\ 
SGR\,J1830$-$0645           & 10.42  & 0.7  & 5.5     & $2.4\times10^{32}$     & 24 & 10$^{\rm{f}}$  &              & X             & \citet{2021ApJ...907L..34C}\\
Swift\,J1818.0$-$1607       & 1.36   & 9    & 3.5     & $1.4\times10^{36}$     & 0.24 & 4.8        &                & $\gamma$ X R& \citet{2020ApJ...896L..30E}\\ 
Swift\,1555.2$-$5402        & 3.86   & 3.05 & 3.47    & 2.09$\times10^{34}$    & 2.01 & 10$^{\rm{f}}$&              & X             & \citet{Enoto-2021}\\
\bottomrule									
\end{tabular}
}
\end{threeparttable}
\begin{list}{}{}\footnotesize{
\item[$^\dagger$] Underwent a giant flare (see section \ref{sec:transients}).
\item[$^{\rm a}$] Assuming a force-free magnetosphere and an aligned rotator, a star radius $R = 10$~km and moment of inertia $I = 10^{45}$~g~cm$^2$, the dipolar component of 
the surface magnetic field at the polar caps is given by $B_{\rm{p, dip}} \sim 2 \cdot (3c^3 IP\dot{P}/8\pi^2 R^6)^{1/2} \sim 6.4 \times 10^{19} (P\dot{P})^{1/2}$~G. Relativistic 
magnetohydrodynamic simulations of pulsar magnetospheres have shown that the estimate offered by this formula is correct within a factor of $\sim2-3$ (Spitkovsky 2006).
\item[$^{\rm b}$] With the same assumptions, the rotational energy loss is given by $\dot{E}_{\rm {rot}} = 4\pi^2 I\dot{P} P^{-3} \sim 3.9 \times 10^{46} \dot{P}P^{-3}$~erg~s$^{-1}$.
\item[$^{\rm c}$] With the same assumptions and assuming that the spin period at birth was much smaller than the current value, the characteristic age is given by $\tau_c=P/2\dot{P}$. 
%\item[$^{\rm d}$] $\tau_k$ indicates the real age, obtained by kinematic measurements. Values are taken from Vigan\`o et al. (2013).
\item[$^{\rm{d}}$] $\gamma=$soft gamma/hard X, X=X-rays, O=optical, I=infrared, R=radio.
\item[$^{\rm e}$] The spin period and its derivative were detected only following the 2008 re-activation of the source. We assume the same spin period derivative also for the 1998 
outburst, and consider the same values for $B_{\rm{p, dip}}$, $\dot{E}_{\rm {rot}}$ and $\tau_c$ in our searches for correlations.
\item[$^{\rm f}$] The value for the distance is assumed.}
\end{list}
\label{tab:magnetars}
\end{table*}
%%%%%%%%%%%%%%%%%%%%%%%%%%%%%%

\section{Magnetars' steady emission} \label{mag:pheno}

\subsection{X-rays and $\gamma$-rays}

Magnetars' persistent X-ray emission is where most of their energy is released. In the 0.5--100\,keV energy range, magnetar spectra are consistently well described by one or two thermal blackbody components, often accompanied by one or two power-law components (see also figure \ref{fig:Xray}). Most magnetar spectra do not reach energies above 10\,keV, with some exceptions that can instead be visible up to hundreds of keV. The X-ray luminosity of magnetars in quiescence ranges between $10^{31-35}$ erg\,$s^{-1}$ and it is believed to originate from thermal processes. The blackbody spectral models have temperatures in the $kT \approx$ 0.15--1\,keV. These temperatures are significantly higher than that of typical rotation-powered pulsars, being powered by the secular decay of their strong magnetic fields in the crust (see section \ref{mag:thermalevo} and figure \ref{fig:mtmodels}). For the same reason, magnetars are usually more luminous in the X-rays than rotation-powered pulsars of similar characteristic age, since their surfaces get continuously heated by the decay of the magnetic field \citep[e.g.][]{aguilera08, vigano13, 2016PNAS, 2020ApJ...903...40D, 2020NatAs.tmp..215I, 2023MNRAS.523.5198D}. Interestingly, the blackbody fitted to the magnetars' X-ray spectra often indicate a thermal emission region much smaller than the stellar surface. This suggests that using a single-temperature blackbody model might be an oversimplification. Strong magnetic fields in the star's crust could lead to substantial temperature anisotropies \citep[e.g.][]{2016PNAS,  2020NatAs.tmp..215I}. These small, highly heated areas could result from particle bombardment by magnetospheric currents, driven by large-scale twists in the external magnetic field or local internal twist kept stable by the internal field helicity.

Regardless of the origin, the thermal emission is likely affected by the presence of a magnetized atmosphere and by magnetospheric processes, such as resonant cyclotron scattering by charged particles. Since the particles are distributed across vast regions of the magnetosphere, which have varying magnetic field strengths, this scattering process generates a broad spectral hardening, rather than narrow lines or distinct harmonics. The two power-laws components present in some objects have photon index within the ranges of $\Gamma_{\rm soft} \sim 2$–4 (in the 0.5--10\,keV soft X-ray energy energy range) and $\Gamma_{\rm hard }\sim$0.5--2 (in the 10--200\,keV hard X-ray energy energy range).
Whereas the soft X-ray non-thermal component was discovered alongside the first magnetar X-ray counterparts, the hard X-ray components have been observed only several decades thereafter, thanks to the advent of sensitive instrument in the 2000's, namely \textit{INTEGRAL}, \textit{Suzaku}, and \textit{NuSTAR} \citep{kuiper04, kuiper06, gotz06, gri07, denhartog08, dkh08, an14, enoto17}. At even higher energies, upper limits derived at MeV, GeV and even TeV energies observations with \textit{CGRO},  
\textit{Fermi}, and \textit{HESS} suggest that these hard power-law components do not persist beyond $\sim500\,$keV \citep[e.g.][]{denhartog08, li17, aleksic13}. They can vary with the pulse phase, and in time when the source is undergoing an outburst, but their luminosity is generally comparable to or even greater than that found below 10\,keV. 
The exact origin of magnetar non-thermal soft and hard X-ray emission, modeled by the two power-laws, is still debated.  However, it is generally thought that they are produced by up-scattering photons off non-relativistic (for the $\Gamma_{\rm soft}$) and ultra-relativistic electrons (for the $\Gamma_{\rm hard}$) \citep{baring07, fernandez07, wadiasingh18} potentially linked to relativistic outflows occurring in the NS magnetosphere \citep{beloborodov13} . 
Note that all these spectral components are necessarily modulated by a substantial interstellar photoelectric absorption in the line of sight, given that all Galactic magnetars lie within the Galactic plane.

The X-ray emission of magnetars is modulated by their spin, showing single, double or even triple peak pulse profiles, depending on the distribution of the X-ray emission region. Moreover, variations in pulse profiles and energy-dependent shifts in the pulse peaks are observed across different phases \citep{dkh08, gri07}. Magnetars' analysis of the evolution of the spin period in time have revealed large timing noise and the presence of glitches in many cases \citep{dib08}. The strong and unstable twisted magnetosphere of these objects leave a chaotic imprint in their timing stability.

%%%%%%%%%%%%%%%%%%%%%%%%%%%%%%%%%%%%%%%%%%
\begin{figure}
    \centering
    \includegraphics[width=0.75\linewidth]{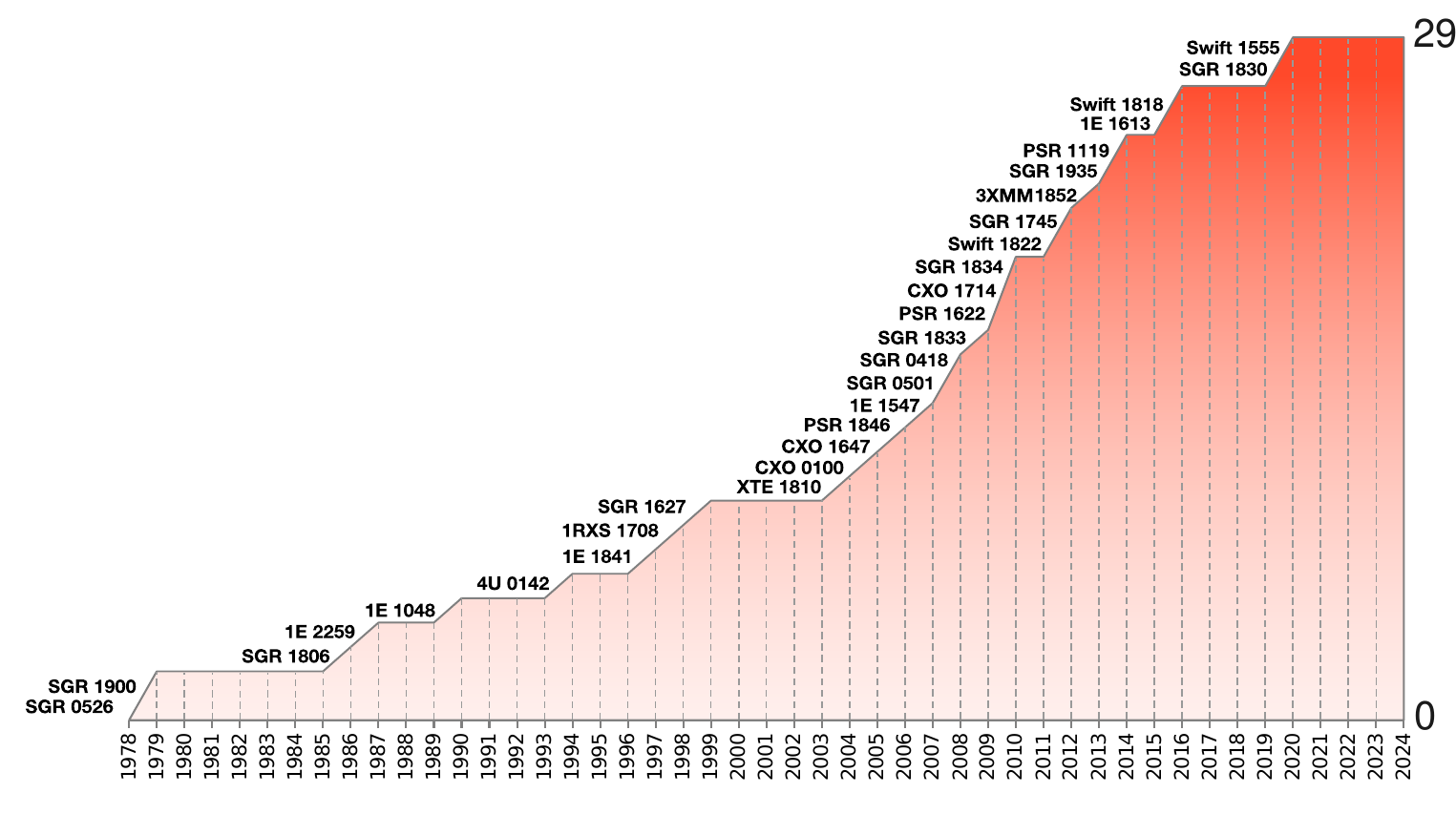}
    \caption{Magnetars discoveries as a function of time. See table \ref{tab:magnetars} for the complete names, which are given here in shortened form. }
    \label{mag:discoveries}
\end{figure}
%%%%%%%%%%%%%%%%%%%%%%%%%%%%%%%%%%%

\subsubsection{X-ray polarization}\label{mag:polar}
The launch of the \textit{IXPE} satellite in 2021 allowed or the first time to perform a long program of polarization measurements in the X-ray band. As polarization is associated to strong magnetism, magnetars were a natural target for this instrument, even though the large number of photon required to resolve spectral and polarization features made so that only the brightest ones could be observed \citep[see the review in][]{2024Galax..12....6T}.

\begin{figure*}
    \includegraphics[width=\textwidth]{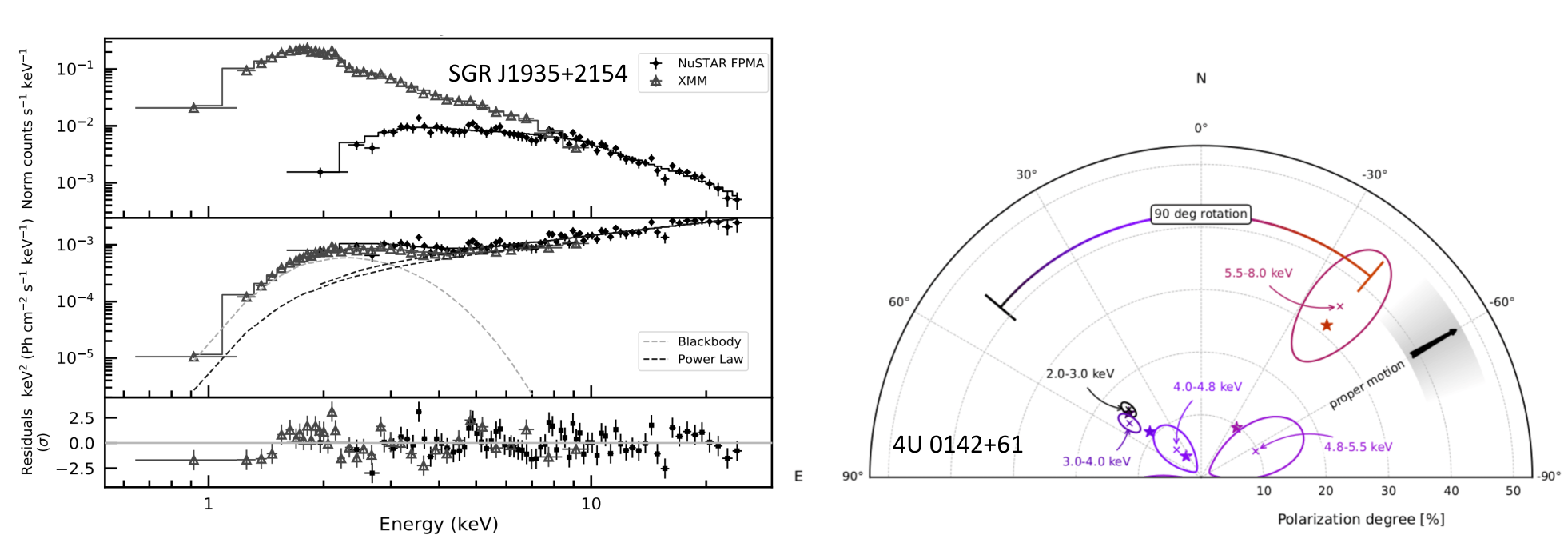}
    \caption{Soft and hard X-ray spectrum of SGR\,1935+2154 \citep{2022MNRAS.516..602B}; X-ray polarimetry detected by \textit{IXPE} for 4U\,0142+61 \citet{2022Sci...378..646T}.}
     \label{fig:Xray}
\end{figure*}

Magnetars were indeed to be strongly polarized, with integrated polarization degrees of $10\%\lesssim PD\lesssim35\%$, reaching up to $\approx80\%$ in confined energy bands (as in the case of 1RXS\,J170849.0-400910, the most strongly polarized source observed by \textit{IXPE} to date). Moreover, in the case of the brightest magnetar, 4U\,0412+61, the polarization angle has been observed to strongly depend on energy, with a $90$\deg swing between low and high energy as shown in Fig. \ref{fig:Xray}; this points towards the fact that different regions on the surface of the magnetar have with rather different optical properties (see section \ref{mag:model_spec}).

%%%%%%%%%%%%%%%%%%%%%%%%%%%%%%%%%%%%
%%%%%%%%%%%%%%%%%%%%%%%%%%%%%%%%%%%%

\subsection{Optical and IR}
Roughly a third of all known magnetars have detected counterparts in the optical or infrared bands (see table \ref{tab:magnetars}), though identifying these is challenging due to their intrinsic faintness ($K\sim21$\,mag, and $V\sim24$ mag) and the Galactic reddening in the disk, where all magnetars lie. During long-term outbursts, the infrared and optical variability of magnetars does not consistently correlate with their X-ray flux, possibly due to limited multi-wavelength observation campaigns, resulting in reports of both correlated and erratic variation patterns \citep{2004A&A...416.1037H,rea04,2004ApJ...617L..53T}. 
Some associations are firmly established through spin modulation in the optical band, seen in sources like 4U\,0142$+$61, 1E\,1048.1$-$5937, and SGR J0501$+$4516 \citep{Kern2002,2011MNRAS.416L..16D}, with additional long-term variability strengthening these associations in other cases.
For 4U\,0142$+$61, a significant dataset recorded with the James Webb telescope, Hubble and the major terrestrial telescopes in the infrared and optical bands, suggests a peculiar spectral decomposition. Some years ago, it was proposed the presence of a multi-temperature disk potentially formed from supernova fallback material and passively heated by the magnetar’s X-rays, aligning with observed X-ray and infrared emission correlations \citep{Hare2024}. However, magnetospheric origin theories also exist, with models suggesting that the curvature radiation in a pair-dominated inner magnetosphere could produce the observed infrared/optical luminosity and explain optical pulsations (see figure \ref{fig:optical}). Recent James Webb observations supports this latter scenario \cite{Hare2024}. The optical and infrared pulsations appear nearly aligned with X-ray profiles, exhibiting broad modulation and substantial pulsed fractions.

\subsection{Radio}
A subset of magnetars also emits in the radio band. Radio emission from magnetars has been discovered in 2006 \citep{camilo06} following the discovery of the first X-ray outburst from these objects \citep[and section \ref{sec:transients}]{2004ApJ...609L..21I}. Radio-emitting magnetars show intermittent or episodic bursts, always coinciding with X-ray outbursts. 

\begin{figure*}
    \includegraphics[width=\textwidth]{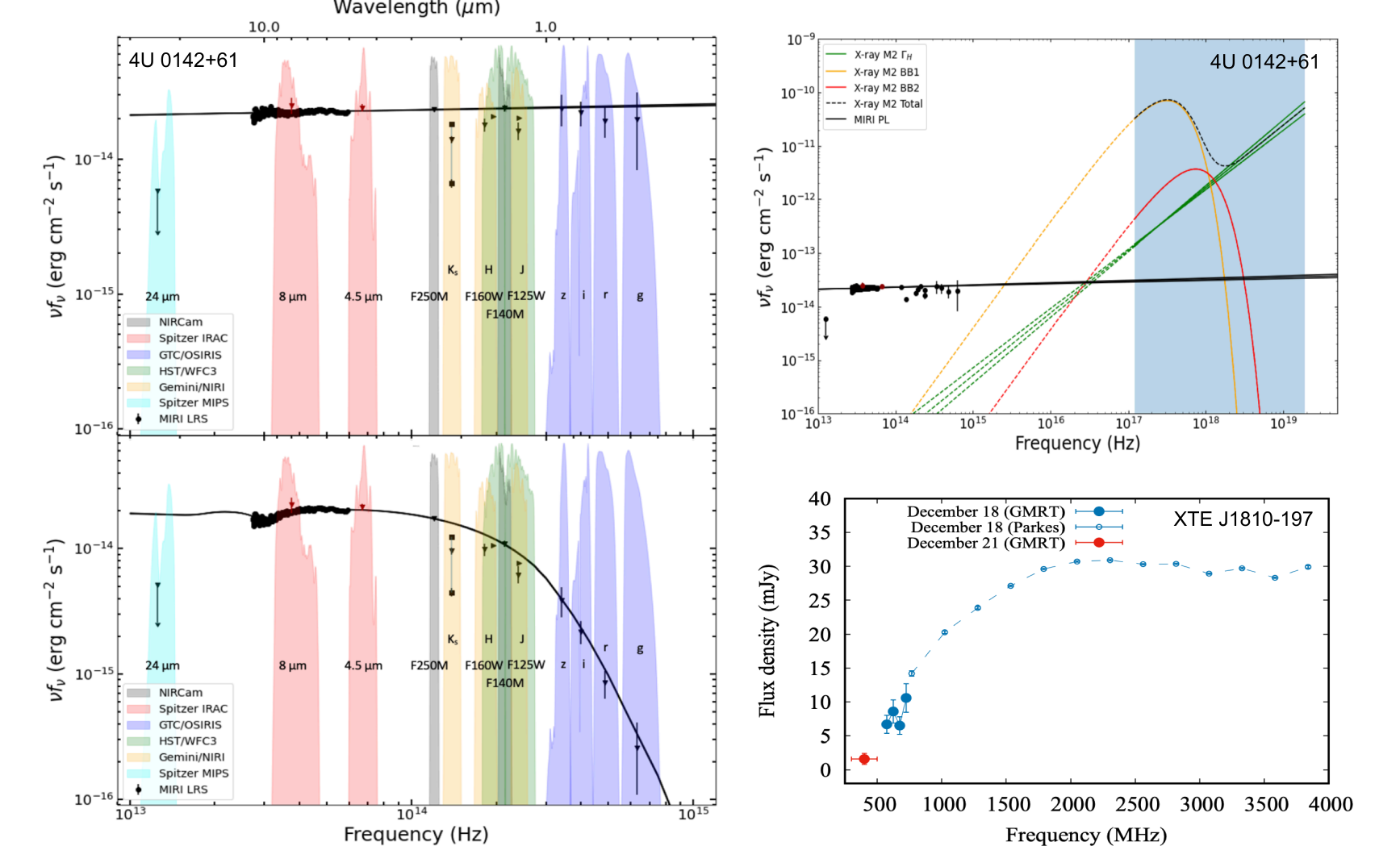}
    \caption{Multi-band spectrum of 4U\,0142$+$61 as observed by the James Webb telescope \citep{Hare2024}. Bottom-right panel show the radio spectrum of XTE\,J1810-197 \citep{Maan2022}.}
     \label{fig:optical}
\end{figure*}

The mechanisms behind magnetar radio emission is still unknown, but the relatively high rotational power of radio magnetars compared to their peers might suggest a similar rotation-powered origin as for traditional pulsars \citep[see also figure \ref{fig:mtmodels}]{rea12}. However, while typical radio pulsars exhibit stable pulse profiles, steep inverted spectra and stable fluxes, the emission profiles of radio magnetars are variable and complex, characterized by highly dynamic and powerful bursty peaks. Unlike the predictable radio signals from pulsars, magnetar radio pulses can show extreme variability in brightness, spectral properties, and polarization. For example, XTE\,J1810$-$197 displays unusual spectral behavior (see figure \ref{fig:optical}), such as rapid shifts in pulse profiles and strong polarization, suggesting that its radio emissions may originate from processes specific to the magnetar's intense magnetic field.

Moreover, radio-emitting magnetars have been observed with very high brightness temperatures, comparable to those of fast radio bursts (FRBs), leading to speculations that magnetars might also be sources of some FRB events. 
Most notably, the Galactic magnetar SGR\,J1935$+$2154 exhibited a burst with FRB-like properties in 2020, providing a compelling link between magnetar activity and FRBs \citep{2020ApJ...898L..29M, 2020Natur.587...54C, 2020Natur.587...59B}. The polarization and temporal structures of these emissions are actively studied, as they offer insights into the geometry and magnetospheric processes of magnetars, which differ markedly from those seen in standard pulsars.

Another trait sometimes found in magnetars with radio emissions is the transition between radio-loud and radio-quiet states, often triggered by energetic events or outbursts (see section \ref{sec:transients}) in higher-energy bands. This transition suggests a close coupling between the magnetar’s magnetic field dynamics and its radio emission capability, once more hinting at unique emission mechanisms directly tied to its powerful magnetic field and the surrounding plasma environment \citep{archibald17}.

%%%%%%%%%%%%%%%%%%%%%%%%%%%%%%%%%%%%
%%%%%%%%%%%%%%%%%%%%%%%%%%%%%%%%%%%%

\section{Magnetar's flaring and outbursts}\label{sec:transients}

\begin{figure*}
    \centering
 \includegraphics[width=0.9\textwidth]{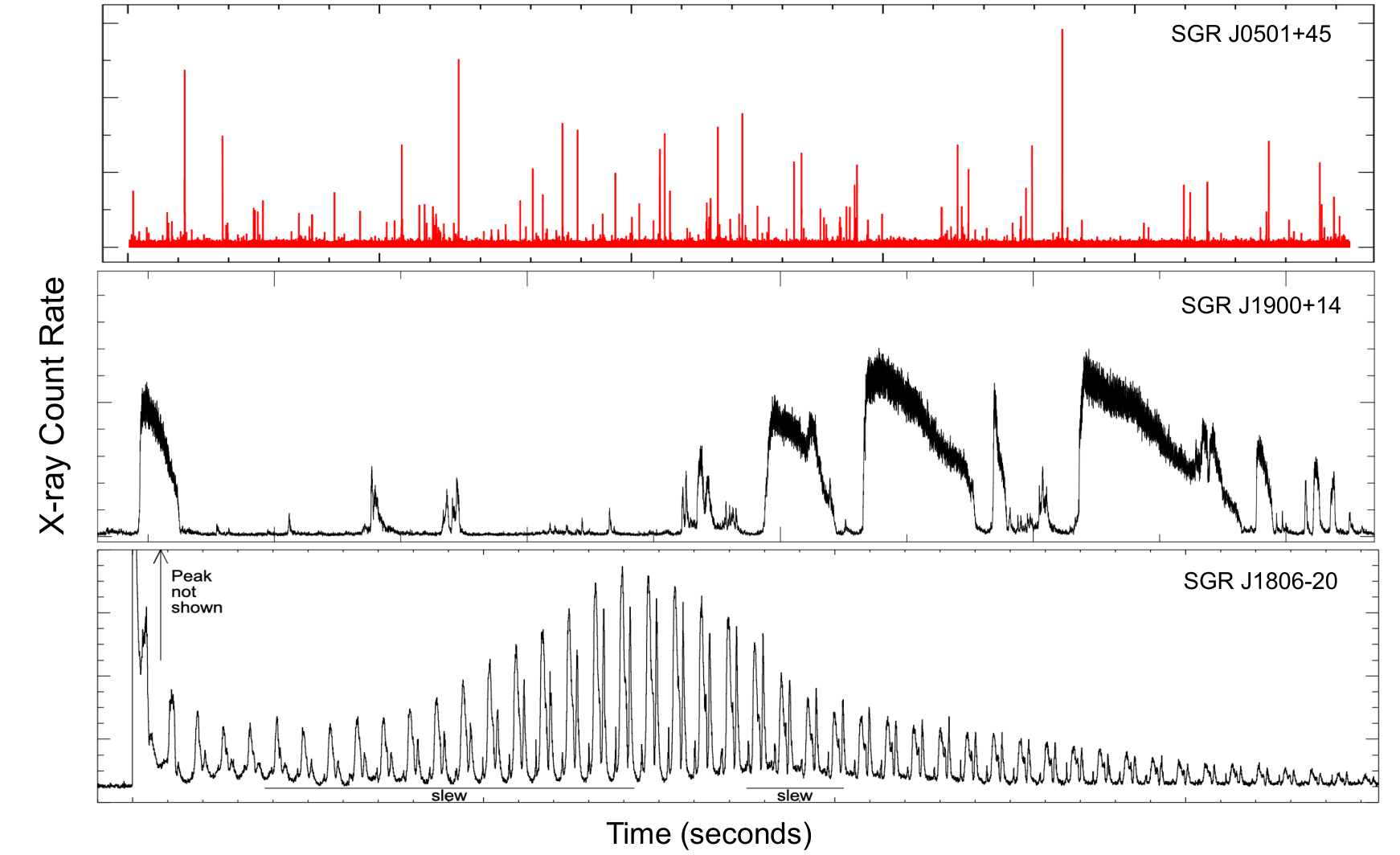}
    %\caption{Bursts}
     \caption{X-ray flaring activity of magnetars (from top to bottom): short X-ray flares \citep{rea09}, intermediate flares \citep{israel08} and a giant flare \citep{palmer05}.}\label{fig:bursts}
\end{figure*}

The flaring activity of magnetars is a distinctive property of these highly magnetic NS, setting them apart from other types due to their unpredictable explosive behavior. Short flaring episodes are often associated with a general increase of the luminosity of these objects, lasting months to years, usually called outbursts. From an observational perspective, magnetars transient activity can fall into three main (strongly related) categories (see also figure \ref{fig:bursts}): \\
\noindent
$\bullet$ {\bf X-ray/$\gamma$-ray Short Bursts}. These are the most frequent and least energetic flares. Short bursts are brief (lasting about 0.1–0.2 seconds), have thermal spectra, and peak at luminosities of $10^{38}-10^{40}$ erg/s, much higher than the Eddington limit for typical NS. Short bursts are irregular in timing, occurring singly or in clusters \citep{kaspi03, woods05}.

\noindent
$\bullet$ {\bf Intermediate Flares}. These events have durations between short bursts and giant flares, ranging from around 1–60 seconds, with luminosities reaching $10^{41}-10^{42}$ erg/s. Some intermediate flares last longer than the magnetar’s spin period, showing clear pulsation at the NS’s spin frequency. Such events are seen in SGRs but have not yet been confirmed in AXPs.

\noindent
$\bullet$ {\bf Giant Flares}. Giant flares are among the most energetic known flaring events in our Galaxy, reaching $10^{43}-10^{45}$ erg/s and rivaling only supernova explosions. Only three have been recorded: in 1979 from SGR\,J0526$-$66, in 1998 from SGR\,1900$+$14, and in 2004 from SGR\,J1806$-$20. Each of these flares featured a short, intense peak lasting under a second, followed by a long, pulsating tail (spanning hundreds of seconds) that aligns with the NS’s spin period \citep[and reference therein]{turolla15}. Moreover, some high energy events in spatial coincidence with a known nearby galaxy have been identified as magnetar giant flares; in particular, recently a giant-flare-like event has been observed in the M82 galaxy, and for the first time a prompt multi-wavelength campaign was able to exclude a GRB origin for it \citep{2024Natur.629...58M}.

\noindent
$\bullet$ {\bf Outbursts}. Transient events are a defining feature of magnetars, acting as key indicators of their presence and contributing significantly to new discoveries in this category of NS. Since the identification of the first transient magnetar event in July 2003 \citep{ibrahim04}, the count of magnetars has grown by a third within six years, largely due to these episodic outbursts \citep[figure \ref{mag:discoveries} and][]{rea11, 2018MNRAS.474..961C}. To date, about 20 magnetar outbursts have been observed, displaying a varied phenomenology, each giving valuable insights into their unique nature.

%%%%%%%%%%%%%%%%%%%%%%%%%%%%%%%%%%
%%%%%%%%%%%%%%%%%%%%%%%%%%%%%%%%%%%

During an outburst, magnetars can exhibit enhanced emissions across multiple bands, from radio to hard X-rays, with soft X-ray fluxes typically increasing by factors of 10 to 1000 relative to their quiescent levels. This flux increase is often accompanied by spectral evolution, where the X-ray spectrum tends to soften as the outburst decays. Decay timescales for these events vary widely among sources, lasting anywhere from a few weeks to several years. The decay profile can follow different patterns, some resembling an exponential decay, others a power-law, and some requiring more complex models with multiple components to describe the flux decrease over time (see figure \ref{fig:outburst_models}).
Recent multi-band observations have advanced our understanding of these outbursts, showing that the emission properties evolve across the spectrum during outburst phases. Furthermore, some outbursts exhibit enhanced radio and optical activity in addition to X-rays, providing clues about the magnetospheric processes and extreme magnetic fields that drive these events \citep{ibrahim04, rea13,younes17, cotizelati18, Borghese2021, Younes2021,Ibrahim2024}.

\section{Modeling Magnetars}

\subsection{Modeling spectra}\label{mag:model_spec}

As mentioned in section \ref{mag:pheno}, the spectrum of a magnetar is  the result of thermal photons emitted from the hot surface that then interact with a complex magnetosphere. The energy scale of the emission is thus determined by the temperature of the star, whose evolution is described in section \ref{mag:thermalevo}; in the following, the processes  behind the shape of the spectrum will be reviewed.

The thermal component arising from the cooling of the NS surface is often modeled in terms a single BB component for simplicity. However, the outermost layers of the surface can have non-trivial optical properties, shaping the spectrum beyond this simple model. In particular, the surface of a magnetar can be covered by a thin gaseous atmosphere, which broadens the spectrum, or lay bare. The latter case is the result of the condensation of the surface induced by the magnetic field \citep{1974IAUS...53..117R}: as the Lorentz force on electrons becomes comparable to or exceeds the electrostatic attraction exerted by the nucleus, atomic orbitals get deformed to an oblong shape, acquiring a different set of quantum numbers with respect to the non-magnetized case. These cylinder-like orbitals can also form covalent bonds between each other, resulting in long chains of atoms bond along field lines (\emph{polymerization}) that then interact to form a solid. The complexity of treating theoretically the properties of this transition, which occurs in a regime entirely inaccessible to ground-based experiments, has produced a range of diverse estimates \citep[e.g.][]{1977ApJ...215..291F,1986MNRAS.222..577J,1987PhRvA..36.4163N,1992AnPhy.216...29F,2006PhRvA..74f2508M}; because of this, the density of the condensed phase of a species with atomic and mass numbers $Z$ and $A$ is expressed within the literature as \citep{2013A&A...550A..43P}
\begin{equation}
    \rho_s=561\,\xi \,AZ^{-0.6} \left(\frac{B}{10^{12}\,\rm{G}}\right)^{1.2} \,\rm{g\, cm}^{-3},
\end{equation} where $\xi\approx1$ is an uncertainty parameter reflecting different estimates. Even more uncertain is the critical temperature of this transition; as an example, \citet{2013A&A...550A..43P} describe the results by \citet{2006PhRvA..74f2508M} as $T_c=5\times10^4 Z^{1/4} (B/10^{12}\,\rm{G})^{3/4}\,$K. At any rate, the emission from a condensed surface is quite different from an atmosphere; from a spectral point of view, it is much more similar to a pure BB \citep[e.g.][]{2004ApJ...603..265T}. Moreover, the two situations differ in their polarization properties, as an atmosphere is expected to strongly polarize radiation (PD$\gtrsim$70\%) mainly in the X-mode (i.e., perpendicularly to the field direction), whereas a condensed surface can only reach PD$\lesssim30\%$ in a direction that varies with the energy and the field orientation \citep[][and references therein, see also section \ref{mag:polar}]{2024Galax..12....6T}.  

Within a strongly magnetized plasma, the electron scattering cross section acquires an additional \emph{resonant cyclotron scattering} (RCS) contribution proportional to \citep[e.g.][]{PhysRevD.3.2303}
\begin{equation}
    \sigma_T\,\frac{\omega^2}{(\omega-\omega_B)^2+\Gamma^2/4}
\end{equation} where $\sigma_T$ is the Thomson cross-section %, $\alpha$ is the angle between the electron momentum and the magnetic field
and $\Gamma=4e^2\omega_B^2/3m_e c^3$ is the natural line width (i.e., the inverse of the characteristic time of the transition between adjacent Landau levels); in astrophysical conditions, this width is generally smaller than the Doppler width $\Delta\omega=\beta_\parallel\omega_B$, where $\beta_\parallel=(k_BT_\parallel/m_ec^2)^{1/2}$ with the temperature $T_\parallel$ being computed over the distribution of momenta parallel to the field \citep{1996ASSL..204.....Z}. This produces an absorption line in the spectrum, which has indeed been detected in some sources \citep[e.g.\ the proton line in the spectrum of the low-B magnetar SGR\,0418+5729][]{2013Natur.500..312T}; however, in the most typical magnetar cases the field structure is so complex that rather than a finite set of well-defined lines the combined results is a power law continuum, superimposed to the initial thermal distribution.

The first attempt to characterize soft-X magnetar spectra in these terms was performed by the 1D model \citep{2006MNRAS.368..690L, rea08}, which despite a number of simplifying assumptions has the advantage of producing a manageable analytical model. In particular, they consider only the Thomson cross section to describe the scattering, which is only adequate in the non relativistic, low field regime. When considering the harder part of the spectrum $\gtrsim10\,$keV, the cross section must take into account relativistic and quantum electrodynamics effects, as well as those of the ultra strong field when $B$ is comparable or above $B_Q$ and electrons can populate different Landau levels.
A similar physical mechanism is invoked to account for the power law emerging in the harder X-rays, above $\sim10\,$keV. Namely, photons populate this energy range when they undergo \emph{resonant inverse Compton scattering} (RICS), in which the interaction of thermal photons with highly energetic electrons in the magnetosphere upscatters them \citep[e.g.][and references therein]{ nobili08,zane09,2021heas.confE..36B}.

%%%%%%%%%%%%%%%%%%%%%%%%%%%%%%%%%%%%%
\begin{figure}
    \centering
    \includegraphics[width=0.5\linewidth]{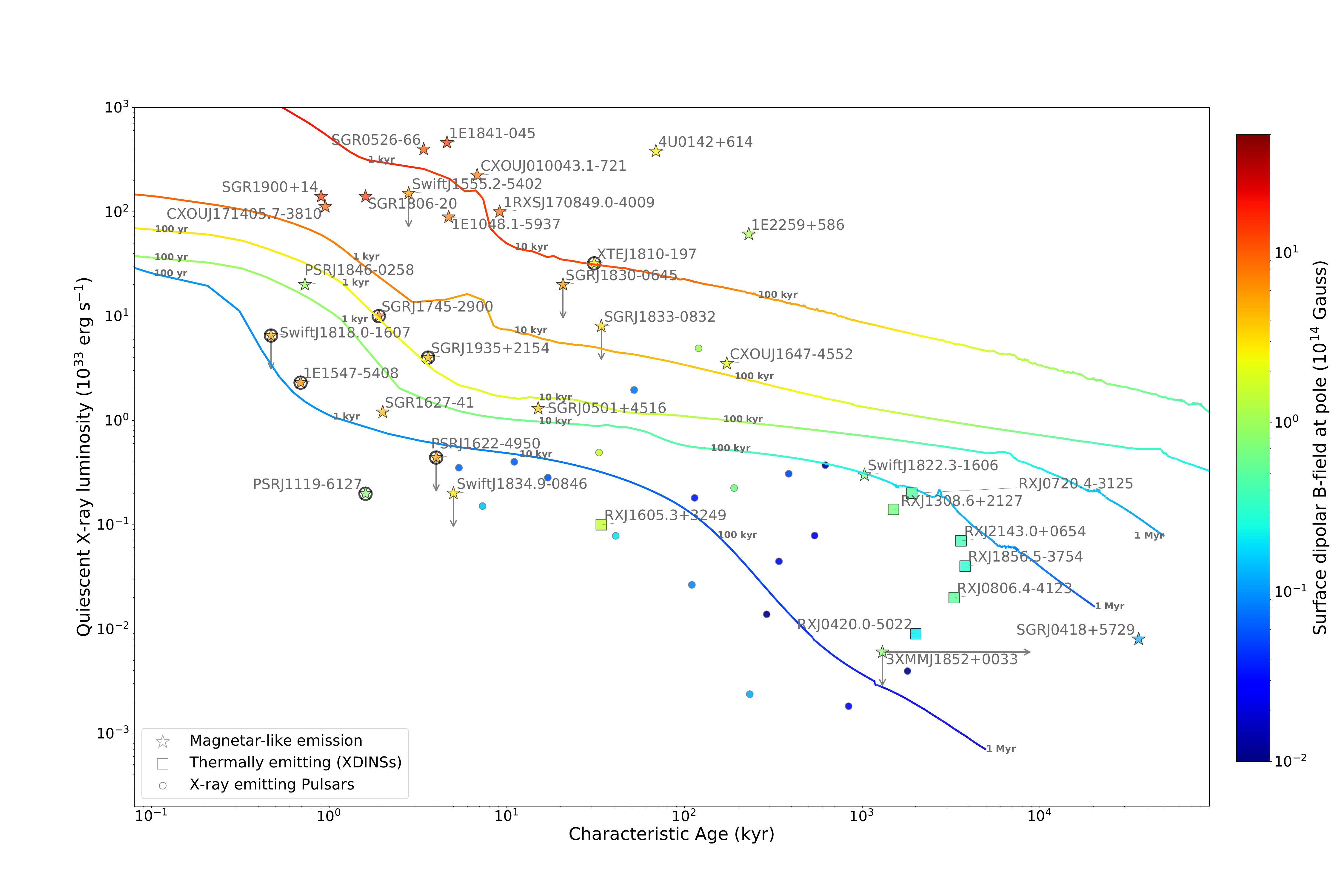}
    \hspace{-0.2cm}      \includegraphics[width=0.5\linewidth]{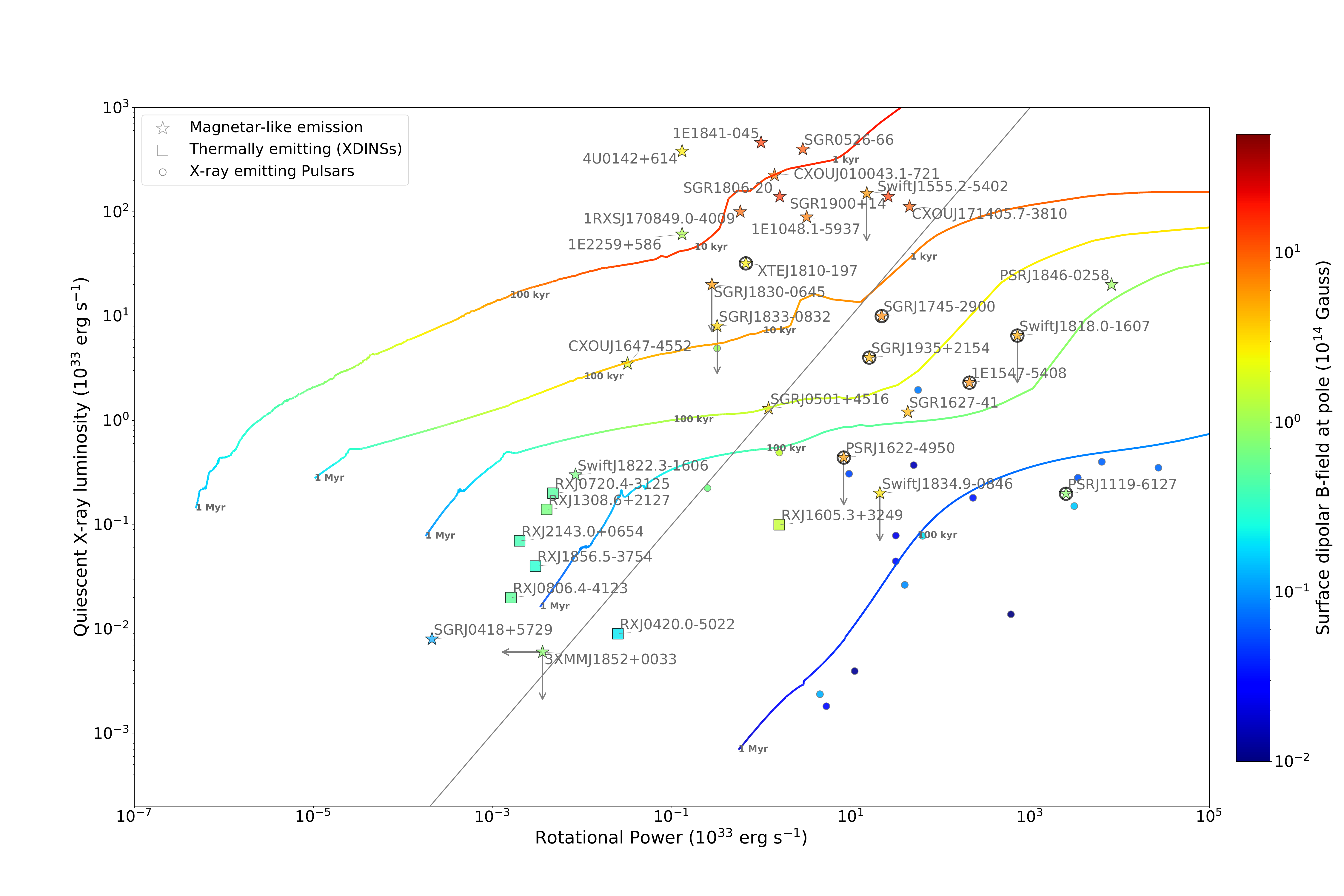}  
    \caption{Thermal X-rays luminosity as a function of the characteristic age $\tau=P/2\dot P$ (left) and rotational power $\dot E\propto\dot P/P^3\,$ (right). The theoretical models are the same as Figure \ref{fig:ppdot}. }
    \label{fig:mtmodels}
\end{figure}
%%%%%%%%%%%%%%%%%%%%%%%%%%%%%%%%%%%%%
\subsection{The evolution of the magnetic field}

The enormous strength of the magnetic field in magnetars very effectively slows down their rotation via braking radiation emission, which is reflected by them lying at the top-right corner of the \ppdot diagram. Still, their periods do not exceed $\sim 10$ seconds. This fact can be accounted for by considering the decay of the magnetic field over the life of the magnetar, and the presence of a highly resistive layer, likely composed of nuclear pasta in the inner crust \citep{2013NatPh...9..431P}.

Under the conditions found in a NS, matter is in general a very good electrical conductor \citep[see][for a review of NS microphysical properties]{transport} but its conductivity is nonetheless finite, hence dissipation is taking place. A crucial aspect of this problem is that the conductivity depends on the temperature and dissipation in turn produces heat, making the magnetic and thermal evolution two strongly coupled problems \citep[e.g.][and references therein]{1996A&A...309..203W,2016PNAS, Pons_2019, 2020ApJ...903...40D, 2020NatAs.tmp..215I, 2023MNRAS.523.5198D, 2024MNRAS.533..201A}.

For what concerns the magnetic field structure, a magnetar can be divided in two main regions: the solid crust and the core. The latter is likely in some kind of superfluid/superconducting state, as indicated by the proneness of magnetars to (anti)glitch \citep[e.g.][and references therein]{2022RPPh...85l6901A}, in which the field gets constrained in flux tubes that have a poorly understood dynamic. More work has been done in the crust, where the bulk of the magnetization is believed to reside; there, the problem can be treated in the eMHD regime, in which the induction equation (neglecting for simplicity GR corrections) reads
\begin{equation}\label{mag:induction}
    \frac{\partial \vec{B}}{\partial t}=-\boldsymbol{\nabla}\times\left[\frac{c^2}{4\pi \sigma}\boldsymbol{\nabla}\times \vec B + \frac{c}{4\pi e n_e} \left(\boldsymbol{\nabla}\times\vec B\right)\times \vec B\right];
\end{equation}
the first term on the RHS describes ohmic dissipation, whereas the second is known as the Hall effect. The latter is non-dissipative, but acts by transferring energy between poloidal and toroidal field components and between different multipoles, most notably in the case of NSs towards the smaller scales in the so-called \emph{Hall cascade}, \citealp{1992ApJ...395..250G}; this can in turn affect dissipation, which is more efficient for the higher multipoles. Moreover, the Hall term is prone to produce MHD instabilities \citep{2014PhPl...21e2110W}, which can dramatically enhance the production of small scale structures \citep{2016PNAS, 2020ApJ...903...40D}. By computing the scale values in to the two terms in Eq.\ \ref{mag:induction}, one can define the timescales for the two processes as
\begin{equation}\begin{aligned}
    \tau_H &= \frac{4\pi n_e e L^2}{c B} \simeq 6.6\times 10^4 \,\left( \frac{n}{10^{34}\,\rm{cm}^{-3}}\right)\left(\frac{B}{10^{13}\,\rm{G}}\right)^{-1}\left(\frac{L}{1\,\rm{km}}\right)^2 \,{\rm{yr}}\\
    \tau_O &= \frac{4\pi \sigma}{c^2}L^2 \simeq 4.4\times10^6\,\left(\frac{\sigma}{10^{24}\,{\rm{s}}^{-1}}\right) \left(\frac{L}{1\,\rm{km}}\right)^2 \,{\rm{yr}}
\end{aligned}\end{equation}
where the numerical values where computed assuming typical values for the crust of a magnetar. The two mechanisms are hence operating over quite different timescales; moreover, the field dissipation time does not explicitly depend from the field itself. In figure \ref{fig:ppdot}, we show a set of NS evolutionary tracks on the \ppdot diagram; in the early phases, they almost follow a line of constant characteristic field, but then progressively diverge, and after $\sim100\,$kyr the field decay is so substantial that there is almost no increase in period. This is indeed reflected by observations, as no magnetars with a period larger than $\sim20\,$s are observed.

\begin{figure}
\centering
    \includegraphics[width=0.4\textwidth]{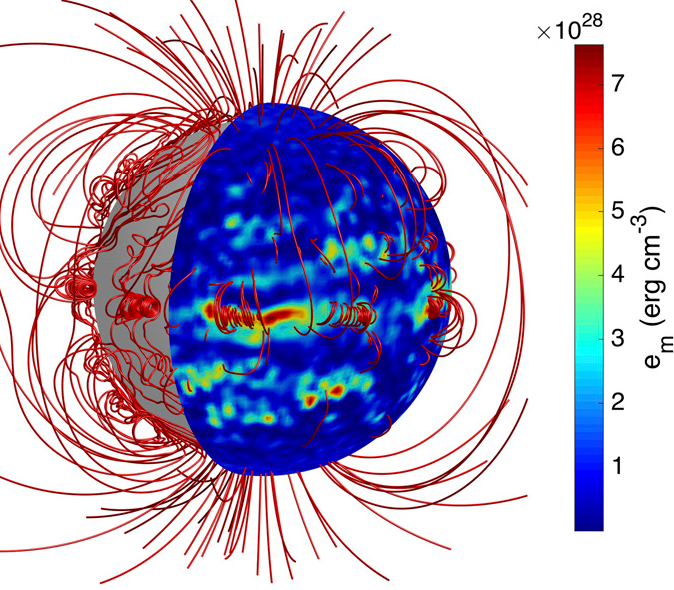}\quad
    \includegraphics[width=0.48\textwidth]{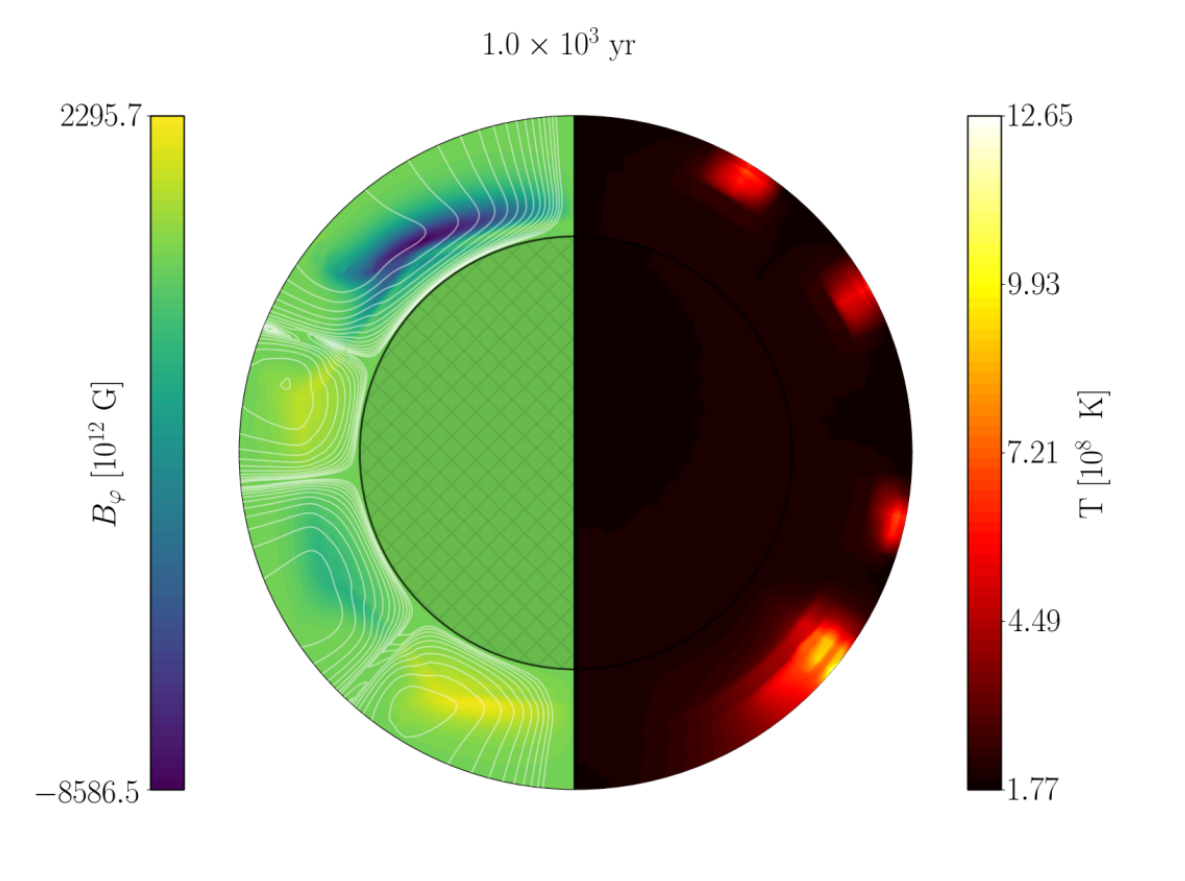}
    \caption{(Left) Magnetic field lines and surface energy after 15\,kyr of the 3D Hall evolution of a field having initially a very small degree of asymmetry ($10^{-4}$ of the total energy in the asymmetric part); from \citet{2016PNAS}. (Right) Example of magnetothermal simulation of a young NS in axial symmetry (the crust is enlarged by a factor 8 for visualization). The coupled evolution leads to the formation of small scale thermal and magnetic features; from \citet{2020ApJ...902L..32D}.  }
    \label{fig:failures}
\end{figure}

A major problem in setting up simulations of magnetic field evolution is represented by the choice of the initial conditions. In fact, it is not clear at present what physical mechanism is responsible for the creation of a magnetar-strength field, and hence what kind of topology is to be expected in a newborn magnetar. In fact, the conservation of the magnetic flux during the core-collapse of the progenitor \citep[\emph{fossil field scenario}, e.g.][]{2006MNRAS.367.1323F} might be able to sizably amplify the magnetic field, but the available data on stellar field do not seem to be able to account for the strong fields found throughout the magnetar population \citep{2021MNRAS.504.5813M}. Another possibility is that the fossil field act as the seed for some kind of dynamo in the proto-NS phase, when the star is spinning extremely fast; still, no consensus on the exact mechanism has been reached, the main candidates being the convective dynamo \citep[as in the original work by][see also \citealp{2003A&A...410L..33B,2022ApJ...924...75M, 2022ApJ...926..111W}]{1993AIPC..280.1085T}, the magnetorotational instability (MRI) driven dynamo \citep[e.g.][]{2009A&A...498..241O,2021A&A...645A.109R,2022A&A...667A..94R} and the Tayler-Spruit dynamo in a PNS spun by fallback material \citep{2022A&A...668A..79B, 2024arXiv240701775B}. More recently, \citet{2024arXiv240808819D} showed that under certain conditions the field evolution described by Eq.\ \ref{mag:induction} may on its own induce an inverse cascade boosting the dipolar field. At any rate, it appears clear that the formation and evolution of ultra-strong fields involves complex topologies, producing structures spanning several scales, as illustrated in figure \ref{fig:failures} (left panel, from \citealp{2016PNAS}, see also \citealp{2022ApJ...936...99D,2023MNRAS.523.5198D}).

Moreover, the magnetic field decay also affects the thermal luminosity of a magnetar. Figure \ref{fig:mtmodels} shows the evolution of the thermal luminosity as computed via the thermal evolution equation, 
\begin{equation}\label{mag:thermalevo}
     c_V\frac{\partial T}{\partial t}=\boldsymbol{\nabla}\cdot\left(\boldsymbol{\kappa}\,\boldsymbol{\nabla} T \right)-Q_\nu+ \left(\frac{c}{4\pi}\right)^2\frac{|\boldsymbol{\nabla}\times \vec{B}|^2}{\sigma}
\end{equation}which is coupled to Eq.\ \ref{mag:induction} not only by the dependence on the magnetic field and temperature of the transport coefficients (heat capacity $c_V$ and conductivity $\kappa$, which is made a $3\times3$ tensor by the magnetic field), but also the last term on the RHS, which describes the Joule heating of the crust (and is the counterpart of the Ohmic dissipation term in the induction equation). Therefore, whereas the thermal evolution of ordinary NSs is mostly one of cooling, regulated by the neutrino emissivity term $Q_\nu$ \citep{2001PhR...354....1Y} and the emission of thermal radiation, in the case of magnetars field dissipation provides a substantial source of additional heating; this is reflected by the higher luminosity the curves in figure \ref{fig:mtmodels} can reach.

\subsection{Modeling transient activity}
As described in section \ref{mag:pheno}, high energy, transient activity is the hallmark feature of magnetars. Owing to the large range of spatial and temporal scales involved, as well as the many different physical processes taking place, the efforts in modeling magnetar transients typically concentrate on either the description of magnetospheric processes or those happening within the star itself, with initial efforts to mathematically couple the two regions so far been able to tackle the long-term evolution only \citep{2023MNRAS.524...32U}; therefore, we will review in the following the two cases separately, even though in reality magnetar activity results from and is influenced by the dynamics in both regimes.

%%%%%%%%%%%%%%%%%%%%%%%%%%%%%%%%%%%%%%%%%%%
%%%%%%%%%%%%%%%%%%%%%%%%%%%%%%%%%%%%%%%%%%%
    
\subsubsection{Transients \& the Crust}
One of the mechanisms often invoked as a trigger for transients is the release of elastic stress built up within the crust as a result the evolution of the magnetic field. The lattice structure of the crust, coupled with the enormous gravitational force, makes it extraordinarily resistant, as calculated by \citet{2010MNRAS.407L..54C}; still, ultra-strong fields, more so if tangled upon small scales by the Hall effect, can overcome the maximum yield. Several studies solving the magneto-thermal evolution in 2 or 3D \citep{2011ApJ...727L..51P, 2020ApJ...902L..32D,2020ApJ...903...40D} and comparing the maximum yield from microphysical calculations to the magnetic stress tensor $M_{ij}=B_iB_j/4\pi$ found that indeed in the early stages of the evolution of a strongly magnetized NS when the crust can fail due to the combined effect of field tangling and local Joule heating. Figure \ref{fig:failures}, right panel, shows an example of an early phase of a magnetothermal simulation in axial symmetry from \citet{2020ApJ...902L..32D}, where the formation of localized hot, magnetized regions is apparent; the authors used these simulations to prove that young magnetars are indeed prone to develop crustal failures.

These violent events are often dubbed \emph{starquakes}, even though the analogy with geophysical processes must be taken with care, as the strong gravity makes so that no cracks and voids are formed, but rather the crustal material starts flowing plastically \citep{2019MNRAS.486.4130L}. The energy thus released can thereafter be converted to heat, possibly via Hall waves \citep{2019ApJ...881...13L}, causing the formation of a hot spot. Field lines moving with the crust they are pinned to may also cause a reconfiguration of the magnetosphere.

\begin{figure}

    \includegraphics[width=\textwidth]{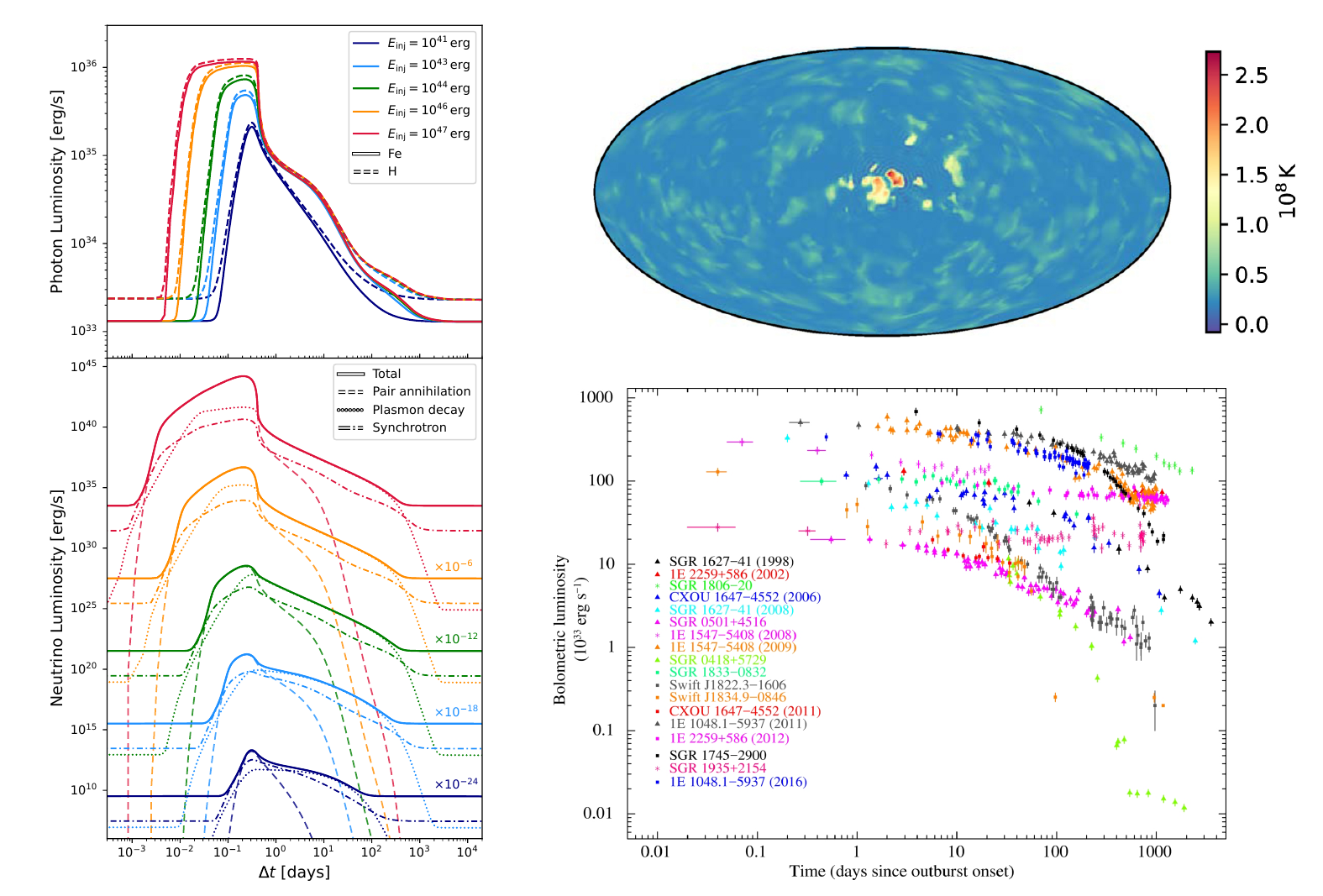}
%    \caption{Outburst simulation....}
    \caption{(Left) Cooling of a 3\,km hot feature in which a heat quantity $H$ has been released over a short amount of time (10\,h) in the outer crust; $\Delta t$ indicates the time since the heating inception. The top panel shows the photon luminosity, in the case of a Fe (solid lines) or H (dashed lines) envelope model; the bottom one the corresponding neutrino luminosity, with its dominant channels (dashes lines: e$^+$e$^-$ pair annihilation; dotted lines; plasmon decay; dot-dashed: weak synchrotron emission); from De Grandis et al., in preparation.
    (Top Right) Surface temperature of a NS with a tangled field during the cooling of a hotspot, which gets broken into several hot features by transport along field lines; from \citet{2022ApJ...936...99D}.
    (Bottom Right) Evolution of the luminosity in the 18 magnetar outbursts recorded before 2018; from \citet{2018MNRAS.474..961C}.
    }
    \label{fig:outburst_models}
\end{figure}

The subsequent study of the cooling of the hotspot can be performed with a machinery akin to that used for secular cooling. This approach was first followed by \citet{2012ApJ...750L...6P}, who found that the cooling timescales indeed match those observed for the evolution of magnetar outbursts, and highlighted the importance of neutrino emission from weak processes in the crust, which gets triggered above $\sim 3\times 10^9\,$K \citep{2001PhR...354....1Y}, in regulating the peak luminosity that an outburst can reach; more detailed simulations (De Grandis et al., in preparation, Fig \ref{fig:outburst_models} top left panel) confirmed that neutrino emission from the outermost layers, in particular electron-positron weak annihilation and plasmon weak decay. This same result was confirmed by \citet{2022ApJ...936...99D} in a 3D setup, which also allowed to address the critical role of the field topology in determining the cooling properties of the outburst. In particular, the presence of a very tangled field makes so that the transport of heat to the source can happen in a complex pattern (figure \ref{fig:outburst_models}, top right panel), suggesting that the values of the effective radii of hotspots as obtained by spectral fits must be regarded as effective ones, describing a more complicated underlying geometry in realistic cases.
    
%%%%%%%%%%%%%%%%%%%%%%%%%%%%%%%%%%
%%%%%%%%%%%%%%%%%%%%%%%%%%%%%%%%%%%

\subsubsection{Transients \& the Magnetosphere}

Due to their fast timescales and largely non-thermal nature, magnetar flares are associated with the field activity in the magnetosphere. In particular, they indicate that the field is not in a purely dipolar configuration (the only component that can be estimated from timing) but rather has a more tangled and variable topology.
Whereas this implies the presence of currents in the magnetosphere in much greater abundance than for normal pulsars \citep{1969ApJ...157..869G}, the field is so strong that the magnetization parameter, i.e.\ the ratio of the magnetic pressure to the energy density of the plasma is still high, $B^2/4\pi\rho c^2\gg 1$. This means that the magnetosphere can be treated to a good degree of approximation as a \emph{force-free} plasma, namely one where\begin{equation}\label{mag:forcefree}
    \left(\nabla\times \vec B\right)\times\vec B = \vec 0.
\end{equation} 

In order to characterize the deviation of a field from a dipole, it is useful to define the \emph{twist} angle associated to a field line as 
\begin{equation}
    \psi=\int_p^q \frac{B_\phi}{B\, r\sin\theta} {\rm{d}} \ell = \int_p^q \frac{B_\phi}{B_\theta\sin\theta}{\rm{d}}\theta~ 
\end{equation}
where the integral is taken along the field line itself, $\vec \ell \parallel\vec B$, between the two points $p$ and $q$ where it touches the surface. Under the assumption of axial symmetry, i.e.\ constant twist $\psi$ for all field lines, eq.\ \ref{mag:forcefree} becomes known as the \emph{Grad-Shafranov} equation \citep[e.g.][and references therein]{2009MNRAS.395..753P}; its solution, labeled by $\psi$, have been studied numerically, finding that stable configurations can be obtained for $\psi\lesssim1$, whereas large twists $\psi\gg1$ make the magnetosphere prone to instability and reconnection \citep{2002ApJ...574.1011U, 2019MNRAS.484L.124C}. The shearing motion of the magnetar crust to which a line is pinned can increase the twist, triggering such unstable regime; this can directly be linked to flaring activity via the emission of Alfvén waves \citep[e.g.][and references therein]{2019ApJ...881...13L}, as well as to outbursts via the heating of the crust by returning currents developing during the untwisting \citep{2009ApJ...703.1044B}.

\section{Magnetars as Engines of Energetic Phenomena in the Universe}

The vast majority of the known magnetars reside in the Milky Way or the nearby Magellanic Clouds. However, the enormous energy budget they store in rotational and magnetic power can in principle power transient events that can be detected from distances of cosmological relevance. 
The large efforts put in studying the transient Universe in the past years, allowed the discovery of a variety of new astrophysical classes and events, many of them proposed to be related with the formation of a magnetar and/or the typical large magnetar flares.  

\noindent
$\bullet$ {\bf Compact binary mergers}. The recent discover of gravitational waves from a double NS binary \citep{2017ApJ...848L..12A, 2017ApJ...848L..13A}  with the first electromagnetic counterpart as a short GRB and its afterglow, pointed to the production of a temporary magnetar as a viable explanation for the observed blue kilonova emission \citep{2018ApJ...856..101M}.

\noindent
$\bullet$ {\bf Fast Radio Bursts (FRBs)}. These are millisecond-long radio flares reaching us from distant Galaxies, which have also been suggested to be either connected to the formation of extra-Galactic magnetars, or being the radio counterparts of their energetic magnetar giant flares \citep{2017ApJ...834..199L, 2017ApJ...843L..26B}. These have gained particular traction after the \textit{INTEGRAL} satellite detected an X-ray burst from the magnetar SGR\,1935+2154 in temporal and spatial coincidence with an FRB \citep{2020ApJ...898L..29M}. This evidence strongly suggests that at least a fraction of the FRB population is indeed linked to magnetars.

\noindent
$\bullet$ {\bf Gamma-Ray Bursts (GRBs)}.
Several lines of evidence have been discovered that point to magnetars as proposed central engines of many long and short GRBs. In particular, their formation appears a good explanation for the GRB long-lived central engine activity \citep{2011MNRAS.413.2031M}, for the multiple bursts of prompt emission in some GRBs \citep{2015JHEAp...7...64B}, the extended emission in short GRBs \citep{2014MNRAS.438..240G}, the X-ray plateau in both short and long GRBs \citep{2013MNRAS.430.1061R}, and the recently discovered class of ultra-long gamma-ray bursts \citep{2015Natur.523..189G}. 

\noindent
$\bullet$ {\bf Super-luminous supernovae (SLSNe).}
Superluminous Supernov\ae\ are a class of particularly energetic SNe, characterized by an optical magnitude above $-21\,$mag, corresponding to $L\gtrsim10^{10}\,$L$_\odot$; this about $10$ times the value for a normal SN\,Ia, and $100$ times the one of a core collapse SN \citep[see e.g.\ the review in][]{2021A&G....62.5.34N}. Evidence is growing that the formation of magnetars powers also some super-luminous supernovae (SLSNe), their spin-down providing a delayed injection of energy \citep{2012MNRAS.426L..76D, 2015Natur.523..189G}.

\section{Conclusions}

Strongly magnetized neutron stars are unique laboratories to understand plasma and matter under extreme field regimes. The advent of X-ray satellites in the past few decades allowed the discovery of about 30 magnetars in our Galaxy. Magnetars are extremely variable sources, showing flares and outbursts on different timescales, and emitting from radio to soft $\gamma$-rays. The study of magnetars is therefore an active and expanding field, in which many mysteries remain to be solved. This chapter is too short to report exhaustively on the extensive past and present research on magnetars, and we acknowledge and apologize for the inevitable biases towards certain specific topics or publications.

\begin{ack}[Acknowledgments]
NR and DDG are supported by the European Research Council (ERC) under the European Union’s Horizon 2020 research and innovation programme (ERC Consolidator Grant ``MAGNESIA'' No.\ 817661) and from grant SGR2021-01269 from the Catalan Government. NR acknowledges support from the ESA Science Faculty Visitor program to ESTEC (funding reference ESA-SCI-E-LE-054), where most of this work has been carried out.
\end{ack}

\bibliographystyle{Harvard}
\bibliography{bibliography}%,reference2}

\end{document}